\documentclass[12pt]{article}
\usepackage{amsmath, amssymb, graphics}
  \usepackage{graphicx}
\newcommand{\mathsym}[1]{{}}

\newcommand{\beq}{\begin{equation}}\newcommand{\eeq}{\end{equation}}
\newcommand{\eqn}{\begin{eqnarray}}\newcommand{\enn}{\end{eqnarray}}

\def\CN{{\cal N}}

\newcommand{\bbibitem}[1]{\bibitem{#1}\marginpar{#1}}

\newcommand{\be}{\begin{equation}}
\newcommand{\ee}{\end{equation}}
\newcommand{\bea}{\begin{eqnarray}}
\newcommand{\eea}{\end{eqnarray}}

\def\Label#1{\label{#1}%
  \smash{\hbox to0pt{\raise1ex\hbox{\tiny[#1]}\hss}}}
\def\noLabels{\let\Label=\label}
\def\nobbibitem{\let\bbibitem=\bibitem}

\begin{document}

\renewcommand{\thepage}{\arabic{page}}
\setcounter{page}{1}
\noLabels % uncomment for final production
\nobbibitem % uncomment for final production

\begin{titlepage}

\begin{flushright}%\vspace{-2cm}
{\small
%{\tt hep-th/yymmnnn} \\
%UPR-T-XXX,
ITFA-2008-19 }
\end{flushright}
%\vspace{12 mm}

\vfil\
%vfil

\begin{center}

{\Large{\bf Statistical Predictions From \\
\vspace{3mm}
Anarchic Field Theory Landscapes
}}
\vfil
\vspace{3mm}

{\bf Vijay Balasubramanian\footnote{e-mail:
vijay@physics.upenn.edu}$^{,ab}$, Jan de Boer\footnote{e-mail:
J.deBoer@uva.nl}$^{,c}$, and Asad Naqvi\footnote{e-mail:
a.naqvi@swansea.ac.uk}$^{,de}$, }
\\

\vspace{8mm}

\bigskip\medskip
\centerline{$^a$ \it David Rittenhouse Laboratories, University of Pennsylvania, Philadelphia, PA 19104, USA}
\smallskip
\centerline{$^b$ \it School of Natural Sciences, Institute for Advanced Study, Princeton, NJ 08540, USA}
\smallskip
\centerline{$^c$ \it Instituut voor Theoretische Fysica, Valckenierstraat 65, 1018XE Amsterdam, The
Netherlands}
\smallskip
\centerline{$^d$ \it  Department of Physics, Swansea University, Singleton Park, Swansea, SA2 8PP, UK}
\smallskip
\centerline{$^e$ \it  Department of Physics, Lahore University of Management and Sciences, Lahore, Pakistan}

\vfil

\end{center}
\setcounter{footnote}{0}
%%%%%%%%%%%%%%%%%%%%%%%%%%%%%%%%%%%%%%%%%%%%%%%%%%%%%%%
%%%%%%%%%%%%%%%%%%%%%%%%%%%%%%%%
\begin{abstract}
\noindent
Consistent coupling of effective field theories with a quantum theory of gravity appears to require bounds on the
the rank of the gauge group and the amount of matter.   We consider landscapes of field theories subject to such
to boundedness constraints.    We argue that appropriately ``coarse-grained" aspects of the randomly chosen
field theory in such landscapes, such as the fraction of gauge groups with ranks in a given range, can be
statistically predictable.      To illustrate our point we show how  the uniform measures on simple classes of $\CN
= 1$ quiver gauge theories  localize in the vicinity of theories with certain typical structures.   Generically, this
approach would predict a high energy theory with very many gauge factors, with the high rank factors largely
decoupled from the low rank factors if we require asymptotic freedom for the latter.
\end{abstract}
%%%%%%%%%%%%%%%%%%%%%%%%%%%%%%%%%%%%%%%%%%%%%%%%%%%%%%%
%%%%%%%%%%%%%%%%%%%%%%%%%%%%%%%%%%
\vspace{0.5in}

\end{titlepage}
\renewcommand{\baselinestretch}{1.05}  %Line spacing
%%%%%%%%%%%%%%%%%%%%%%%%%%%%%%%%%%%%%%%%%%%%%%%%%%%%%%%
%%%%%%%%%%%%%%%%%%%%%%%%%
%%%%%%%%%%%%%%%%%%%%%%%%%%%%%%%%%%%%%%%%%%%%%%%%%%%%%%%
%%%%%%%%%%%%%%%%%%%%%%%%%%%%%%%%%%%%
\tableofcontents

%\newpage

\section{Introduction}

It is commonly supposed that the huge numbers of vacua that can
arise from different compactifications of string theory
\cite{boussopolchinski, KKLT} imply a complete loss of
predictability of low energy physics.   If this is the case,  the
stringiness simply constrains the possible dynamics rather than
the precise complement of forces and matter.    Every string
theory leads to some effective field theory at a high scale
$\Lambda$, taken to be, say,  an order of magnitude below the
string scale.   Predictions for low energy physics have to made in
terms of this effective field theory.    Thus, the landscape of
string theory vacua leads to a landscape of effective field
theories at the scale $\Lambda$ .     Here we ask if constraints
of finiteness imposed on this landscape via its origin in string
theory might be sufficient to lead to a degree of predictability,
at least in some statistical sense.      Previous authors have
discussed how continuous parameters can scan in a random landscape
of effective field theories \cite{hall, mcallister,
nimaetal,anarchy, nelsonstrassler, hoggartnielsen,gibbons}, and
there has been some study of the gauge groups and matter content
attainable from specific string theoretic scenarios \cite{dienes,
douglas1,douglas2,dijkstra,timoetal,douglastaylor}.  For example,
\cite{timoetal} and \cite{douglastaylor} discuss the distribution
of gauge groups arising in intersecting brane models on torus
orientifolds.

We will impose the weakest of the constraints arising from string
theory -- namely that it should be possible to couple the
effective field theory consistently to a quantum theory of
gravity.  It has been argued \cite
{swampland,swampland2,swampland3} that such consistency with
string theory requires that the rank of the gauge group and the
number of matter fields be bounded from above.\footnote{A possible
bound on the number of matter species in theories containing
gravity was originally discussed by Bekenstein \cite{bekenstein}.}
Since we will not impose any constraints based on rules arising
from symmetry or dynamics on the measure, we will call this an
``anarchic''  landscape, in recollection of the terminology in
\cite{anarchy}.   Thus we will study simple anarchic landscapes of
field theories  bounded in this way,  and illustrate how
statistics  can lead to characteristic predictions for the low
energy physics.      These predictions are strongest  for
appropriately coarse-grained attributes of a theory that possess
the property of {\it typicality} in such landscapes -- i.e. they
are overwhelmingly likely to lie close to certain typical values.
An example of such a typical property will be the fraction of
gauge groups with ranks lying within some range.   We will
illustrate and develop our thinking using some simple examples.

\section{The set of field theories}

A natural, large class of field theories to consider is the set of
quiver gauge theories. For simplicity, we will restrict attention
to $\CN=1$ supersymmetric quiver gauge theories
 where the gauge group is a product
of unitary groups,
\be
G = \prod_{i=1}^L U(N_i).
\ee
In addition, there will be $A_{ii}$ hypermultiplets transforming
in the adjoint of $U(N_i)$, and $A_{ij}$ hypermultiplets
transforming in the $(\mathbf{N}_i,\bar{\mathbf{N}}_j)$ of
$U(N_i)\times U(N_j)$. The nonnegative integer matrix
\be
A_{ij}\geq 0, \quad i,j=1\ldots L
\ee
describes the number of arrows from site $i$ to site $j$ in the
quiver.

Of course, to specify the full gauge theory we also need to
describe the K\"ahler potential for the hypermultiplets, the gauge
kinetic terms, the superpotential and possibly Fayet-Iliopoulos
terms.  We will postpone a discussion of these quantities for now and
will in this paper only   discuss the matter and gauge group
content of the $\CN=1$ theory.

Gauge theories of quiver type are ubiquitous in string theory, and
this is the main motivation to restrict attention to this class.  Bifundamentals tend to appear in string theory
because strings have two endpoints
only.  A typical setup to engineer quiver $N=1$ theories is to
consider D6-branes wrapping 3-cyles inside a Calabi-Yau manifold
in type IIA string theory, in which case the number of
bifundamentals is related to the intersection number of the
3-cycles. By including orientifolds, we can also easily engineer
quiver gauge theories with $SO$ and $Sp$ gauge factors, but we
will postpone a study of these theories to another occasion.

Our goal will thus be to study random $U(N)$ quiver gauge
theories. Before looking at some concrete examples, we are first
going to make some general remarks on possible
further restrictions on the set of gauge theories, on the choice of measure on the space of theories, and the
kinds of properties we might predict.

\subsection{Interesting classes of quiver gauge theories}

Three possible restricted sets of gauge theories are:
\begin{enumerate}
\item \underline{Anomaly free theories}. A simple physical
requirement that we should impose is the condition that
there are no anomalies. This translates into the statement that
for all $i$
\be \sum_{j\neq i} (A_{ij}-A_{ji}) N_j =0. \label{anomaly}
\ee
The expectation value of the left hand side of this equation in
the unconstrained set of quiver gauge theories with the uniform measure
 is clearly zero, as the measure is invariant under
$A_{ij} \leftrightarrow A_{ji}$. Therefore, ``on average,'' random
quiver gauge theories are anomaly free, and one might be inclined
to not worry about anomalies anymore. However, from a physical
point of view it seems rather peculiar to allow forbidden theories
in an ensemble, as it is not at all clear that properties of the
set of anomaly free theories are properly reproduced by the full
set of random quiver gauge theories. Hence we will for the most
part restrict to field theories which obey (\ref{anomaly}).
\item \underline{Asymptotically free theories}. Another natural
constraint we can impose is that the theories we consider are
asymptotically free, which makes them well-defined in the UV.
Asymptotic freedom is less compelling than anomaly cancellation,
as the set of random quiver theories may well represent a set of
low-energy effective field theories obtained e.g. in string
theory. Gauge group factors that are IR free and strongly coupled
in the UV will typically start to act as global symmetries at
sufficiently low energies and will not directly lead to contradictions.
The condition for asymptotic freedom is that for all $i$,
\be \label{free}
A_{ii} N_i +\sum_{j\neq i} (A_{ij}+A_{ji})N_j < 3 N_i.
\ee
This tends to constrain the $A_{ij}$ to not be very large but to
be of order unity instead.
\item \underline{Purely chiral theories.}
If we imagine our field theory to be some effective field theory
at a high scale $M$, assume there are no other dimensionful
parameters around, and write down the most general superpotential,
it will contain many mass terms with masses of order $M$. At
energies below $M$, it makes sense to integrate out all massive
fields with masses of order $M$. The remaining gauge theory will
have fewer fields and will no longer allow for mass terms: all fields
that can be integrated out have been removed. The remaining
set of purely chiral theories with
\be \label{chiral}
A_{ii}=0, \qquad A_{ij}=0\,\,\,{\rm or} \,\,\, A_{ji}=0 \,\,\,
{\rm for} \,\, i\neq j
\ee
are therefore a natural starting point for viewing random quivers
as low-energy effective field theories. Such chiral theories
allow for general cubic superpotentials at the marginal level.  Higher order terms are suppressed by a mass
scale in the Lagrangian, although some quartic superpotentials can become marginal in the infrared.

\item \underline{Equal rank theories.}
In order to simplify the analysis, we could take the
ranks of all the gauge groups to be fixed and equal. For such
theories both the anomaly cancellation constraint as well as the
asymptotic freedom constraint are much easier to implement.
However, we do not have an obvious physical motivation that would
prefer these theories, so they are mainly helpful to develop
intuition for the more general case.

\end{enumerate}

\subsection{Averages and typicality}

Given a  set of gauge theories with a suitable measure on
them, we can compute expectation values of various quantities,
such as the average rank of a gauge group, the average number of
matter fields, etc. Though averages are useful to compute,
they are especially interesting when they also represent the
{\it typical} value of a quantity. Typicality is a notion that exists in situations when a thermodynamic limit can be
taken wherein some parameter $N$, controlling the size of the ensemble, can be taken to infinity.
Then, a quantity enjoys the property of typicality if its probability distribution  becomes more and more narrowly
peaked around its expectation value as $N \to \infty$:
\be
\lim_{N\rightarrow \infty} \frac{ \langle {\cal O}^2 \rangle -
\langle {\cal O} \rangle^2 }{ \langle {\cal O} \rangle^2} =0.
\Label{typical}
\ee
In other words, quantities that are typical  are equal to their
ensemble averages with probability one in the limit $N \to
\infty$\footnote{This criterion is not very useful when $\langle
{\cal O} \rangle=0$. More generally, we should normalize the
operator ${\cal O}$ in such a way that the range of values it can
take is independent of $N$ and then require that the variance
vanishes in the large $N$ limit.}.

Familiar examples of typical operators are statistical mechanical
quantities such as pressure and free energy. Also, we note that
for a standard Boltzmann distribution, for one particular
occupation number with
\be
\langle N \rangle = \frac{\sum_{k\geq 0} k e^{-\beta
k}}{\sum_{k\geq 0} e^{-\beta k}} =
\frac{e^{-\beta}}{1-e^{-\beta}},\qquad \langle N^2 \rangle =
\frac{\sum_{k\geq 0} k^2 e^{-\beta k}}{\sum_{k\geq 0} e^{-\beta
k}} = \frac{e^{-\beta}(1+e^{-\beta})}{(1-e^{-\beta})^2}
\ee
the variance to mean squared ratio appearing   in (\ref{typical})
equals $e^{\beta}$. In other words, a microscopic quantity like a
single occupation number will not be typical.   Observables that
achieve typicality are inevitably coarse-grained -- e.g. the
number of Boltzmann particles with energies between $c/\beta$ and
$(c+\epsilon)/\beta$ for constants $c$ and $\epsilon$ will be
typical.   In studying the statistics of effective field theories
we should be interested in finding appropriately
``coarse-grained'' structures that are typical.

\subsection{Choice of measure}

In order to define and discuss averages and typicality for random
quiver gauge theories, we need to define a suitable measure on
this space. One could imagine that dynamics gives rise to an
interesting and complicated measure. For example, one could
imagine weighing field theories by the dimension or even size of
the cohomology of their respective moduli spaces, having the close
connection between quiver gauge theories and D-brane moduli spaces
in mind.   As another simple example of how dynamics can affect
the measure, if we suppose that dynamical effects can give the
matter fields any expectation value, then generically all the
gauge groups will be broken to $U(1)$ and analysis of the
distribution of gauge factors is moot.   However, in $N=1$
theories of the kind we study, the potential for the matter fields
typically develops isolated minima and the gauge group is broken
to a product of Abelian and non-Abelian factors (for instance, a
cubic superpotential for an adjoint superfield classically breaks
$U(N)\rightarrow U(p)\times U(N-p)$ for some $p$.). Classically,
in the context of Calabi-Yau compactification, one imagines that
the manifold has some set of distinct but intersecting cycles and
the non-abelian factors in the gauge theory are related to the
number of branes wrapped on each cycle. Then strong gauge dynamics
might break these gauge factors further.  For the present we will
ignore such dynamical issues and use a uniform measure subject to
various constraints of boundedness.  Since we are ignoring rules
arising from the underlying dynamics, we will call our measures
``anarchic''.

Finally, in the context of e.g. string landscape discussions, one
might want to associate various kinds of Bayesian measures to
different types of field theories.   For example, to correctly
make statistical predictions for the UV field theory, given our
hypothetical bound on the the matter and gauge groups, we strictly
speaking condition our probability distribution on the known facts
about infrared physics.   From this perspective, we actually want
the uniform measure on a bounded space of gauge theories that,
when run to the infrared, contain the standard model as a sector.
Conditioning in this way, is well beyond our ability at present,
and  so we will simply investigate the uniform measure on bounded
spaces of quiver gauge theories, to study whether and how
typicality occurs.

Experience in statistical physics has shown that directly computing averages and variances over bounded
configuration spaces can be difficult.     Thus, to simplify analysis we can try to use a grand canonical ensemble
to constrain the total rank and the total number of matter fields.   This involves summing over theories with
arbitrary ranks and amounts of matter while including in the measure a Boltzmann factor for the rank of the
gauge group, and a separate Boltzmann factor for the total number of matter fields
\be \label{measure}
\rho \sim \exp(-\beta \sum_i N_i - \lambda \sum_{ij} A_{ij} N_i
N_j ).
\ee
One could also include Boltzmann factors for, e.g., the total number
of nodes, the total number of gauge bosons, etc., but for our purposes (\ref{measure}) will be sufficient to
illustrate the main ideas.    Such an approach only works if the ensemble of theories does not grow
exponentially fast in the total rank and number of matter fields.  If such exponential growth occurs, the
Boltzmann weight does not fall quickly enough for the microcanonical ensemble to be well approximated by the
canonical ensemble.  We will see that the space of theories typically grows too fast with the number of fields to
permit use of the canonical approach to make statistical predictions from a bounded landscape of effective field
theories.

\section{Typicality in Toy Landscapes}

\subsection{Theories without matter: coarse graining and typicality}

As an example of our approach, consider a landscape of field theories with no matter, where the rank of the
gauge group is equal to a large number $N$.  For simplicity, let the gauge group be a product of unitary factors
\be
G = \prod_{i=1}^L U(N_i) \, .
\ee
Then the rank of $G$ is $\sum_i N_i = N$;  thus the $N_i$ form an  integer partition of $N$.    To study the
distribution of gauge factors in an anarchic landscape of such field theories, we can construct the canonical
partition function
\be
Z = \sum_{\{ r_k \}} e^{-\beta \sum_k k \, r_k - \alpha \sum_k
r_k} = \prod_k {1 \over 1 - e^{-\beta \, k - \alpha}}
 \equiv \prod_k {1 \over 1 - u \, q^k}
\Label{nomatterpartition}
\ee
Here $r_k$ is the number of gauge factors of rank $k$, $\beta$ is
a Lagrange multiplier constraining the total rank to be $N$ and
$\alpha$ is a Lagrange multiplier that can be used to constrain
the number of gauge factors; sometimes it is more convenient to
work with $q=e^{-\beta}$ and $u=e^{-\alpha}$ instead. In writing
this we have used measure that treats the ordering of gauge
factors as irrelevant. So, for example, $U (2) \times U(3) \times
U(2)$ is the same as $U(3) \times U(2) \times U(2)$ and so on. In
such a measure, all $U (N_i)$ factors are treated as identical,
and not distinguished from each other by parameters like their
gauge couplings. This measure will be modified if the gauge theory
is realized by wrapping D-branes on cycles of Calabi-Yau because
in that case the locations of branes and the sizes of the cycles
will allow us to distinguish between many different configurations
that lead to the same gauge group. Nevertheless, the present
measure is interesting from a purely field theoretic point of
view, i.e. if one is simply counting field theories, and is
illustrative.

To fix  $\beta$ and $\alpha$ we require that
\be
N = \sum_{j=1}^\infty {j \, u \, q^j \over 1 - u \, q^j} ~~~~ ; ~~~~
L = \sum_j {u \, q^j \over 1 - u \, q^j} \, ,
\Label{NandL}
\ee
where $N$ is the total rank and $L$ is the total number of gauge factors.
We will take
\be
u \sim O(1)  ~~~~;~~~~  \beta \sim {1 \over \sqrt{N}} \,
\Label{nomatterbeta}
\ee
which, we will see later, implies $L \sim \sqrt{N}$.  Then from (\ref{nomatterpartition}) it is easy to show that:
\be
\langle r_j \rangle = {u \, q^j \over 1 - u \, q^j} ~~~;~~~
{\rm Var}(r_j) = {u \, q^j \over (1 - u \, q^j)^2} = { \langle r_j  \rangle \over 1 - u \, q^j} \, .
\ee
The variance to mean squared ratio is
\be
{\rm{Var}(r_j) \over \langle r_j \rangle^2}
=
{1 \over u \, q^j} = e^{\beta j + \alpha} \geq e^{\alpha} \geq O(1) \, .
\ee
To get the last inequality we simply used $\alpha, \beta > 0$.
Thus we see that in a universe with such anarchic landscapes, the
number of gauge factors $r_j$ with rank $j$ is not typical in the
sense defined in (\ref{typical}) and thus cannot be predicted with
confidence.

However, we could ask whether there are any more coarse grained
structures in such landscapes which are more predictable. For
example, consider the number of gauge factors whose ranks lie
between $c \sqrt{N}$ and $(c + \epsilon) \sqrt{N}$ where $c$ and
$\epsilon$ are both $O(1)$:
\be
\langle R(c,\epsilon) \rangle \approx
\int_{c \sqrt{N}}^{(c + \epsilon) \sqrt{N}} dj \, \langle r_j \rangle = {1 \over \beta}
\ln \left[ {1 - u \, e^{-(c + \epsilon) \sqrt{N} \beta} \over 1 - u \, e^{-c \sqrt{N} \beta}} \right] \, ,
\ee
where we approximated the sum as an integral.  The variance of this coarse-grained variable is
\be
{\rm Var}(R(c,\epsilon)) =
\int_{c \sqrt{N}}^{(c + \epsilon) \sqrt{N}} dj \, {\rm Var}(r_j)
= {u \over \beta} \left[ {e^{-c \sqrt{N} \beta} - e^{- (c + \epsilon) \sqrt{N} \beta} \over
(1 - u \, e^{- c \sqrt{N} \beta}) ( 1 - u \, e^{-(c + \epsilon) \sqrt{N} \beta})}\right] \, ,
\ee
where used the fact that in this canonical ensemble the $r_j$ are statistically independent variables.  Thus, for $
\beta \sim 1 / \sqrt{N}$ (\ref{nomatterbeta}),
\be
\langle R(c,\epsilon) \rangle \sim O(\sqrt{N}) ~~~~;~~~~
{\rm Var}(R(c,\epsilon)) \sim O(\sqrt{N}) ~~~~\Longrightarrow~~~~
{\rm{Var}(R(c,\epsilon)) \over \langle R(c,\epsilon) \rangle^2} \sim O(1 / \sqrt{N}) \,.
\ee
This means that the variance to mean squared ratio vanishes in the large $N$ limit indicating that $R(c,\epsilon)
$ is a typical variable.   Thus, in such anarchic landscapes, the number of gauge factors with ranks between $c
\sqrt{N}$ and $(c + \epsilon) \sqrt{N}$ can be predicted with high confidence.   Also,  approximating the second
equation in (\ref{NandL}) as an integral, the total number of gauge factors turns out to be
\be
L =  - {\ln(1 - u) \over u \beta} \sim O(\sqrt{N}) \, .
\ee
By the above variance analysis this number will also be typical.  Thus, in such anarchic landscapes, the total
number of gauge factors is highly predictable.    These results follow essentially because the unordered
partitions of a large integer enjoy a sort of central limit theorem  -- representing such partitions by a Young
diagram, one can show that in the large $N$ limit, the boundary of an appropriately rescaled diagram
approaches a limit shape encoded by the $\langle r_j \rangle$ computed above \cite{vershik}.

\subsection{Cyclic, chiral quivers}

Above we saw how suitably coarse-grained aspects of the structure of a randomly chosen field theory in a
bounded landscape might be statistically predictable.   The next step is to add matter to the theory to see how
this changes the analysis.     As we discussed, we must insist that matter is added in an anomaly-free way and
implementing this constraint is one of the main difficulties in studying studying statistical ensembles of quiver
gauge theories.   Thus, to make a beginning,  we will study cyclic, chiral quiver gauge theories for which
anomaly cancelation is very easy to implement.

In cyclic quivers, each gauge group is connected to the next one
by bifundamentals, with the circle being completed when the last
group connects to the first one.      Taking the ith group around
the circle to be $U(N_i) $, the constraint on the total rank will
be $\sum_i N_i = N$.   So, as in the example without matter, the
number $N_i$ form a partition of $N$.   Anomaly cancellation
requires that each gauge group have as many fundamentals as
antifundamentals.   It is easy to show that,  the minimal solution
to anomaly cancellation constraints is that the number of
bifundamentals between $U(N_i)$ and $U(N_{i+1})$ is
\be
A_{i (i +1)} =  C^{-1} \cdot \prod_{l\neq i,(i+1)} N_l ~~~~;~~~~ C
= \rm{GCD}(\{ \prod_{l\neq i,(i+1)} N_l \}) \Label{cyclicanomaly}
\ee
All other solutions to the anomaly cancellation equations are
integer multiples of (\ref{cyclicanomaly}).   We will examine an
ensemble in which the matter fields in the gauge theory are
presumed to satisfy (\ref {cyclicanomaly}) in such a way that the
total number of fields comes as close as possible to some bound
$K$. Thus for this setup the matter fields are uniquely chosen
once the gauge groups are selected.   (More generally, we could
have imagined an ensemble where the number of matter fields was
allowed to vary, in which one would need to sum over multiples of
$A_{i(i+i)}$ subject to a bound.  This is difficult to do here
since the  GCD of the products of integer subsets appearing in the
denominator of (\ref{cyclicanomaly}) is presumably sporadically
behaved.)

One key difference from the matter-free case, is that the {\it
order} in which the gauge groups appear on the ring of the quiver
is important.  In general, different orderings will lead to
different quiver gauge theories, except when the permutations
correspond to symmetries of the quiver, such as the cyclic
permutations of the nodes, or cyclic permutations combined with
one reflection.    These are just elements of the dihedral group
corresponding to the symmetries of a regular polygon with vertices
corresponding to the nodes of the quivers. Additional symmetries
will arise if some of the $N_i$ are equal and we will treat the
exchange of groups with identical ranks as giving the same theory.
This sort of measure would arise if we imagined our field theory
landscape as arising from D-branes on a Calabi-Yau in which all
the cycles give rise to gauge theories with the same coupling,
which could for example happen if we would resolve an $A_k$
singularity in such a way that all two-cycles would have equal
size.

\subsubsection{The canonical ensemble breaks down}
We will first try to analyze the statistics of cyclic, chiral
quivers in a canonical ensemble. All along, as motivated above, we
will assume that the gauge groups uniquely fix the matter content.
Let $r_k$ be the number of times the group $U(k)$ appears. Then,
the total rank $N$, and the number gauge factors $L$, are
\be
N= \sum_k k r_k   ~~~~;~~~~ L = \sum_{k} r_k \, .
\ee
We want to compute the partition function of this ensemble of {\it ordered} partitions of $N$:
\be
Z= \sum_{\{r_k \} }   {1 \over 2} \Bigl(\sum_k r_k -1\Bigr)!  ~e^{ -\beta \sum_k k r_k  - \alpha \sum_k  r_k }  \prod_k
{1 \over r_k!}  \, .
\Label{Zcan}
\ee
The combinatorial factor that appears here is simply the number of
ways we can choose $r_1,r_2,\ldots$ gauge group factors out of
$\sum_k r_k$, divided by $2(\sum_k r_k)$ to account for the cyclic
and reflection symmetry of the quiver\footnote{This counting
actually ignores certain accidental symmetries that can arise in
the structure of the quiver.    For example, in a cyclic quiver in
which the gauge groups $U(N_1)$ and $U(N_2)$ alternate, only one
cyclic permutation gives a different configuration for the quiver.
The fully correct counting can be derived using Polya theory -- we
are simply using the leading terms of that analysis, which is
sufficient for our purposes.}. Rewriting the partition function in
terms of the $ \Gamma$ function, we obtain
\be
Z
 =  {1\over 2} \sum_{\{r_k \} }  \Gamma\Bigl(\sum_k r_k \Bigr)  \prod_k   ~{e^{-\beta  k r_k - \alpha r_k}\over r_k!}
 \Label{gamma}
\ee
Using the integral representation of the $\Gamma$ function
\be
\Gamma(z) =
\int_0^\infty dt ~t^{z-1} e^{-t}
\ee
the partition function can be rewritten as
\be
Z = {1 \over 2} \int_0^\infty dt~{ e^{-t}  \over t} ~ \sum_{\{r_k \} }   \prod_k   { t^{r_k}  \over r_k!} e^{-\beta  k r_k -
\alpha r_k }
\ee
Exchanging the sum and the product, and after some manipulations, we obtain
\be
Z =  {1 \over 2} \int_0^\infty dt~{ e^{-t}  \over t} ~  \exp\Bigl( {t e^{-\alpha} e^{-\beta} \over 1-{e^{-\beta}}}\Bigr) \, .
\ee
This integral is only convergent if
\be
{e^{-\alpha} e^{-\beta} \over 1-{e^{-\beta}}}  < 1 ~~~\Rightarrow ~~~ e^{-\beta}  < { 1\over 1+e^{-\alpha}} \equiv e^
{-\beta_H}\, .
\Label{limitbeta}
\ee
This implies that there is a limiting $\beta$ above which the partition function is undefined, because the
integrand diverges as $t \rightarrow \infty$.  There is also always a divergence as $t \rightarrow 0$ which  can
be regulated by recognizing that the divergence is a constant independent of $\alpha$ and $\beta$.  To show
this, we define $\gamma = {e^{-\alpha} e^{-\beta} \over 1-{e^{-\beta}}}$, and find that
\be
{d Z \over d \gamma} =   \int_0^\infty dt~{ e^{-(1-\gamma)t} }  = { 1 \over 1 - \gamma}
\ee
which implies that, below the limiting temperature,
\be
Z = - \log ( 1 - \gamma) = - \log( 1 - {e^{-\alpha} e^{-\beta} \over 1-{e^{-\beta}}}) = -\log( 1 - {uq \over 1-q})
\ee
where $u= e^{-\alpha}$ and $q = { e^{-\beta}}$.

In order to achieve a large total rank, $\beta$ must be tuned to
close to its limiting value $\beta_H$ (\ref{limitbeta}).   Then,
if, for example, we put $u=1$, the expectation value of the total
rank is
\be \langle N \rangle = q\frac{\partial}{\partial q} \log Z \sim
\frac{-1}{2\epsilon \log (4\epsilon)}
\ee
where we tuned $q=q_H-\epsilon=\frac{1}{2}-\epsilon$ to get a large rank.   Similarly, in this approximation we
can compute
\be
\langle r_k \rangle \sim \left( \frac{1}{2} \right)^{k+1}
\frac{-1}{2\epsilon \log(4\epsilon)} \sim \left( \frac{1}{2}
\right)^{k+1} \langle N \rangle \, .
\Label{canonicalav}
\ee
This is completely different from the matter-free result for the typical partition: for example, on average one
quarter of the nodes will be abelian.   However, we also find that
\be
{\rm Var}(r_k) \sim \left( \frac{1}{2} \right)^{2r+2}
\frac{-1}{(2\epsilon)^2 \log(4\epsilon)} \sim - (1+ \log(4\epsilon)) \, \langle r_k \rangle^2
\Label{canonicalvar}
\ee
This is much larger (as $\epsilon \rightarrow 0$) then the expectation value squared.   In other words, the
number of group factors with a given rank is not typical in the sense of (\ref{typical}).

As in the matter-free case, we might wonder if a more
coarse-grained question would have a more statistically
predictable answer.  For example, we might ask how many gauge
factors we expect to see within some range of ranks.    The mean
and variance of such a coarse grained variable can be extracted by
summing over the quantities in
(\ref{canonicalav},\ref{canonicalvar}) because the $r_k$ are
independent random variables in our ensemble.    In the central
limit theorem, summing $M$ identically distributed random
variables reduces their fluctuations because both the mean and the
variance are enhanced by a factor of $M$; thus   the variance to
mean squared ratio is {\it reduced} by a factor of $M$.   In the
matter-free example, something like this happened because,
although the $r_k$ were not identically distributed, their
dependence on $k$ was sufficiently to weak to allow the central
limit theorem to work.  In the present case, the exponential
dependence of (\ref{canonicalav}, \ref{canonicalvar}) on the rank
$k$ means that this mechanism fails -- the mean and the variance
remain dominated by the smallest $k$ in the range of the sum.
Thus, it would appear that there is no simple statistically
predictable quantity in this landscape.

However,  this is in fact happening because the canonical ensemble
is breaking down and is not a good approximation of the
microcanonical ensemble anymore.    The canonical ensemble will
reproduce the microcanonical ensemble when the growth of the
configuration space with the total rank is slow enough so that,
when multiplied by an exponential Boltzmann factor, a nicely
localized measure results. Here the Gamma function and the
exponential in (\ref{gamma}) compete on a equal footing and lead
to a widely spread out measure in which the rank of the gauge
group fluctuates wildly over the ensemble, leading to a very large
variance.   Indeed, we should expect this sort of behavior to
occur generally when studying the statistics of quivers since the
number of graphs increases rapidly with the number of nodes.  Thus
we turn to the microcanonical ensemble in order to implement more
accurately our constraint on the total rank.

\subsubsection{Microcanonical analysis}
We consider once more a cyclic quiver and ignore accidental
symmetries.    The microcanonical partition function for cyclic
gauge theories of rank $N$ and $L $ nodes is simply the number of
such theories.  This is given by the coefficient of $q^N$ in
\be
\frac{1}{2 L} \left[ q + q^2 + q^3 + \ldots \right]^L \, .
\ee
Here the $1/2L$ divides out the cyclic permutations and reflections.  We find that
\be
Z_L=\frac{1}{2 L} \left( \begin{array}{c} N-1 \\ N-L \end{array}
\right).
\ee
Summing this over $L$ we can write a partition function which is
canonical in the number of nodes and microcanonical in the total
rank $N$:
\be
Z(u) = \sum_{L=1}^N u^L \, Z_ L =
{(1+u)^N -1 \over 2 N} \, .
\ee
To get the unbiased landscape in which all theories of equal rank
have equal weight, we should take $u=1$, but we will consider
other values of $u$ as well. The expectation value of $L$ is
\be
\langle L \rangle = u  \partial_u \log(Z(u)) ={ u (1+u)^{N-1}  \over (1+u)^N -1 }~ N \, .
\ee
For the unbiased ensemble with $u=1$, we get $\langle L \rangle=
{N \over 2}$ in the large $N$ limit. However, if $u \sim {1 \over
\sqrt{N}}$, then $\langle L \rangle \sim \sqrt{N}$, and if $u \sim
{ 1 \over N}$, then $ \langle L \rangle \sim O(1)$. In fact, if $u
\sim N^{-a}$, $\langle L \rangle \sim N^{1-a}$. It can be checked
that the canonical analysis gives the same expectation values.
However, the microcanonical variance in $L$ is
\be
{\rm Var}(L)  = \Bigl(1 - {Nu \over (1+u)^N -1}\Bigr) { 1\over 1+u} \langle L \rangle
\ee
For the three scalings of $u$, i.e. $u \sim N^{-a}$ , the variance in $L$ is some order 1 number times the mean
value of $L$, independent of $a$.   Thus, when $\langle L \rangle$ is large, the variance to mean squared ratio
is small, unlike the canonical analysis.    This means that in such landscapes  the number of gauge factors is
typical in the sense of (\ref{typical}) and is therefore highly predictable.

The expectation value for the number of abelian factors is:
\be
\langle r_1 \rangle =  {1 \over Z} \sum_{L=1}^N { u^L \over L} L  \left( \begin{array}{c} N-2
L-2 \end{array} \right)
 =  {u^2 (1+u)^{N-2}\over (1+u)^N-1} N
\ee
When $u=1$, this becomes $\langle r_1 \rangle = {1 \over 4} N$ in the large $N$ limit. When $u \sim {1 \over
\sqrt{N}}$, $\langle r_1 \rangle \sim O(1)$. And when $u \sim {1 \over N}$, $\langle r_1 \rangle \rightarrow 0$. In
fact, for $u \sim {N^{-a}}$, $\langle r_1 \rangle \sim N^{1-2a}$.  It can be checked that these expectation values
match the canonical ensemble.  However, the  the variance in $r_1$ is much smaller in the microcanonical
ensemble.  First we compute that
\bea
\langle r_1^2 \rangle& =& {1 \over Z} \sum_{L=1}^N { u^L \over L} \bigg\{ L  \left( \begin{array}{c} N-2 \\
L-2 \end{array} \right) + L(L-1)   \left( \begin{array}{c} N-3 \\
L-3 \end{array} \right) \bigg\}\\
& = & {u^2 ( 1+u)^{N-4} \bigl( u(uN+4) +1 \bigr) \over (1+u)^N -1} ~ N
\eea
Therefore, the ratio of the variance to the mean squared is
\be
{\langle r_1^2 \rangle - \langle r_1 \rangle ^2 \over \langle r_1 \rangle^2} =  { 1 + u \Bigl(4- {Nu \over (1+u)^N-1}
\Bigr) \over (1+u)^2} \times {1 \over \langle r_1 \rangle}
\ee
The coefficient of $1/\langle r_1 \rangle$ in this expression is of $O(1)$ for for $u \sim N^{-a}$, with $0 \leq a \leq
1$.

Pulling everything together, in the unbiased ensemble with $u=1$,  the average number of gauge factors is $N/
2$ and the number of abelian factors is $N/4$.   Both of these quantities are typical in the sense of (\ref{typical})
and hence highly predictable in this landscape without any coarse-graining.   In a biased ensemble with $u \sim
{1 / \sqrt{N}}$, the total number of gauge factors is $O(\sqrt{N})$, and the number of abelian factors is $O(1)$.
Since variance is of the same order as the mean for both quantities, the number of gauge factors is typical and
thus predictable, but the number of abelian factors is not.   In this case, we expect that a coarse-grained statistic,
such the fraction of gauge groups in a given range, would be more predictable as in the matter-free case.

\paragraph{Higher Ranks}
To find the expectation value of the occupation number of rank $r$, we can insert a ``chemical potential'' for that
rank. So
\be
Z(u,\{y_k\}) = \sum_{L=1}^N {u^L \over 2L}  \left[ \sum_{k=1}^N
q^k y_k  \right]^L  |_{q^N}  \, ,
\ee
where the left hand side  equals the coefficient of $q^N$ in the right hand side.
The expectation value $\langle r_k  \rangle $ is given by
\bea
\langle r_k \rangle &=& \partial_{y_k} \log \Bigl(Z(u,\{y_k\})\Bigr) |_{\{y_k\}=1} \\
% & = & {1 \over Z[u]} \sum_{L=1}^N {u^L} q^r  \left[ \sum_{r=1}^N q^r y_r  \right]^{L-1}  |_{q^{N-r}} \\
& = & {1 \over Z[u]} \sum_{L=1}^N u^L    \left( \begin{array}{c} N-k-1 \\
L-2 \end{array} \right)  \\
&=& {u^2 (1+u)^{N-k-1} \over (1+u)^N -1}N
\eea
In the unbiased ensemble with $u\sim 1$, $\langle r_k \rangle \sim (1/2)^{k+1} N$ as we found in the canonical
ensemble.    Similarly, the expectation value $\langle N_r^2 \rangle  $ is:
\bea
\langle r_k^2 \rangle &=&  { 1 \over Z} y_k \partial_{y_k}  y_k \partial_{y_k}  \Bigl(Z(u,\{y_k\})\Bigr) |_{\{y_k\}=1} \\
%& = & { 1\over Z} \Bigl( y_r \partial_{y_r} Z  + y_r^2 \partial^2_{y_r} Z \Bigr)|_{\{y_r\}=1} \\
& = & {1 \over Z}   \sum_{L=1}^N u^L  \bigg\{  \left( \begin{array}{c} N-r-1 \\
L-2 \end{array} \right)  + (L-1) \left( \begin{array}{c} N-2r-1 \\
L-3 \end{array} \right)  \bigg\} \\
& = & {u^2 (1+u)^{N-2r-2} \Bigl( 2u  + (N-2r+1) u^2 + (1+u)^{r+1} \Bigr) \over (1+u)^N-1} N
\eea

So the ratio of the variance to the mean squared is
\be
{ {\rm Var}(r_k) \over \langle r_k \rangle^2 } = {1 \over (1+u)^{k+1}} \bigg\{
(1+u)^{k+1} + u\bigl((1-2k)u+2\bigr) -{Nu^2 \over (1+u)^N-1} \bigg\} \times {1 \over \langle r_k \rangle}
\ee
This is always $O(1)$ times $1 / \langle r_k \rangle$, and hence the number of gauge groups of a given rank is
typical, and hence highly predictable, if the average is large.

\paragraph{Lessons: }   We are finding that in an anarchic landscape of cyclic quiver gauge theories, the actual
number of gauge factors of a given rank is highly predictable.  Specifically, the distribution of ranks is
exponential and the low rank populations are statistically predictable with high confidence.   In a  biased
landscape in which the measure favors having a number of gauge factors that is sufficiently smaller than the
total rank, we found that the number of factors with a fixed rank in not typical in general although the total
number of factors can be.  In this case, one could test whether an appropriately coarse grained quantity, like the
fraction of gauge groups with ranks in some range, is more predictable.

\section{Thinking about the general quiver}

To extend our analysis to the general quiver gauge theory we could try to compute a partition sum of the form
\be
Z = \sum_L \sum_{N_i, A_{ij}}
\exp(-\beta \sum_i N_i - \lambda \sum_{ij} A_{ij} N_i N_j )
\ee
where $L$ is the number of nodes of the quiver, $N_i$ are the
ranks of the gauge groups, and $A_{ij}$ are the numbers of
bifundamentals between nodes $i$ and $j$.  One difficulty is that
this partition sum is canonical and, as we found,  it may not
implement the constraints on the total rank and the amount  of
matter very well because of the rapid growth of the space of
theories.  Secondly the sum should only be over anomaly cancelled
theories.  Thirdly, there are discrete symmetries which tend to
lead to vanishing expectation values.  In view of this, below we
will develop some approaches to dealing with the two latter
issues.

\subsection{Implementing anomaly cancellation}

\paragraph{A loop basis for anomaly free theories: }
If all the gauge groups have the same rank, the general anomaly
free theory can  be constructed by making sure that the
bifundamental fields always form closed loops.     One can always
construct such matter distributions by saying that each of the
possible loops  in the quiver has $n_i$ fields running around it.
Where loops overlap the matter content will either add or subtract
depending on the orientation of the loops (again here we are
supposing that non-chiral doublets decouple; in addition, we
identify negative $A_{ij}$ with a positive $A_{ji}$ and vice
versa.). Any loop in the quiver can be constructed by summation of
a basis of independent 3-loops and it can be shown that this basis
will have
\be
N_L = {(L - 1) (L - 2) \over 2}
\Label{loopcount}
\ee
elements.    For example, consider the case with $L=6$ nodes, i.e.
there are six gauge groups that we label from 1 to 6. Then, the
following three loops form a basis for all loops: (123), (124),
(125), (126), (234), (235), (236), (345), (346), (456). The basis
has 10 elements which is equal to $N_6 = (6-1) (6-2)/2$. We can
check that the $N_L$ loops provide enough free parameters to
parameterize the space of anomaly free theories.    To see this,
note that the solutions to the anomaly cancellation equations form
a vector space of dimension
\begin{equation}
 {L(L -1) \over 2} - (L-1) = {(L-1) (L-2) \over 2} = N_L
 \Label{anomcount}
\end{equation}
where $L(L-1)/2$ is the number of parameters $A_{ij}$ from which
we have subtracted the $(L-1)$ anomaly cancellation conditions on
$L$ groups.

Even when the ranks are unequal, anomaly free theories can be
constructed from a basis of 3-loops because (\ref{loopcount}) and
(\ref{anomcount}) are equal.  However, the links of any given
3-loop will have to be populated with a different number of fields
in a way related to the GCDs of the three groups appearing in it.
For example, suppose one has the three gauge groups $SU(r_1 \cdot
g) \times SU(r_2 \cdot g) \times SU(r_3 \cdot g)$ where $r_i$ are
a triple of positive integers that do not share a common factor
and $g$ is any another positive integer.   Then if we take number
of chiral bifundamentals between gauge group $i$ and $j$ to be
$A_{ij} = \epsilon^{ijk} r_k$, we get an anomaly free
theory.\footnote{As a specific example consider a four-node quiver
with gauge group $SU(3)_1 \times SU(5)_2 \times SU(7)_3 \times
SU(8)_4 $. Then, we can get an anomaly free theory by making a
loop of four with $A_{12}= 7 \cdot 8$,  $A_{23}= 3 \cdot 8$,
$A_{34}= 3 \cdot 5$, $A_{41}= 7 \cdot 5$. We can obtain this as a
sum of two 3 loops: (124)+ (234). The loop (123) corresponds to an
$SU(3)_a \times SU(5)_b \times SU(8)_c$ with $A_{ab}=8$,
$A_{bc}=3$, $A_{ca}=5$ while the loop (234) corresponds to
$SU(5)_i \times SU(7)_j \times SU(8)_k$ with $A_{ij}=8$ ,
$A_{jk}=5$, $A_{ki}=7$. To get the four loop (1234), we need to
cancel the (24) link which means that need to add $7 \cdot (124) +
3 \cdot (234)$. This precisely reproduces the four loop numbers.
Another anomaly free theory could be generated by adding (124) to
(234). In this case, the fields along link (24) will not cancel,
but by construction, the number of fields going into and out of
each gauge group will cancel. }

This way of thinking suggests that one way to do the statistics of
anomaly free theories is to first select a basis of anomaly free
3-loops and then do the statistics of populations of these loops
given a bound on the total number of loops.
%This is hinting at
%the usefulness of some theory of homology for these loops.

\paragraph{Anomaly free, asymptotically free, chiral, equal rank
gauge theories: } This set of theories is very easy to analyze, as there are can be
only five different types of vertices in such quivers (see
Fig.~\ref{fivevertex}). Therefore the most general quiver
arises by combining these five vertices in various combinations.
Superficially, the second vertex with two separate lines coming in
and two separate lines going out allows for the largest amount of
combinatorial freedom and will quite likely dominate this set of
theories. It would be interesting to explore this class further.
Possibly it can be mapped to an existing solvable lattice model in
statistical mechanics.

\begin{figure}
\begin{center}
\includegraphics[scale=0.9]{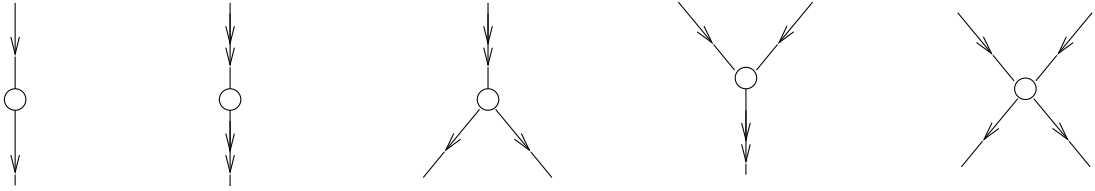}
\caption{The five vertices from which all anomaly-free, chiral,
asymptotically free, equal rank theories are constructed}
\Label{fivevertex}
\end{center}
\end{figure}

\paragraph{Anomaly cancellation for a general quiver by using an extra node: }
If we drop the constraint of asymptotic freedom, the set of  anomaly free, chiral, and equal rank theories
is easy to parametrize.   It is not difficult to see that if we take any set of edges ${\cal
S}$ such that the edges together form a connected tree which
contain all vertices of the quiver, then the anomaly equations
uniquely determine the $A_{ij},A_{ji}$ with $(ij)\in {\cal S}$ in
terms of the $A_{ij},A_{ji}$ with $(ij) \not\in {\cal S}$. Thus we
can simply take an arbitrary set of chiral matter fields for all
edges not in ${\cal S}$, after which anomaly cancellation uniquely
fixes the remaining links.

A simple example of the set ${\cal S}$ is the star-shaped tree
consisting of all edges $(1i)$, $i=2\ldots L$. In other words, if
we remove one vertex and all its edges, and arbitrarily specify
the chiral matter content in the remaining quiver with $L-1$
vertices, this uniquely determines an anomaly free , chiral, equal
rank quiver gauge theory with $L$ gauge groups.  To illustrate
this, consider a four-node quiver. Take $A_{12}=a$, $A_{32}=b$ and
$A_{13}=c$\,\mbox{}\footnote{If we say $A_{12}=a$, then we mean
that $A_{12}=a$ for $a\geq0$ and $A_{21}=-a$ for $a\leq 0$. This
will guarantee that the theory is chiral}. Then anomaly
cancellation uniquely fixes
\be
A_{24}=a+b, \quad A_{41}=a+c,\quad A_{43}=b-c .
\ee

This method can be extended to theories where the gauge groups
have unequal ranks.   First consider an arbitrary chiral, quiver
with $L - 1$ nodes.    Let the rank of the group at the ith node
be $N_i$.  For anomaly cancellation, the net number of
fundamentals minus antifundamentals at each node must be zero.
Let $K_i$ be the net excess matter (number of fundamentals minus
anti-fundamentals) at each node.   Then we can always add an
additional $U(1)$ gauge group with $N_i \, K_i$ bifundamental
fields under this $U(1)$ and the $U(N_i)$ of the ith node. This
will give an anomaly free theory.    This extra node can be
non-abelian, but its rank is restricted to be a divisor of the set
$\{ N_i \, K_i \}$.    In this way, the statistics of general
anomaly free quivers on $L$ nodes can be studied by first
constructing arbitrary $L-1$ node quivers and then adding a extra
node according to the above algorithm.

\subsection{Dealing with discrete quiver symmetries: an example}

From above, the set of anomaly free, chiral and equal rank theories with four nodes is
parametrized by the rank $N$ of the gauge groups and three
integers $a,b,c$. The measure (\ref{measure}) becomes
\be
\rho = \exp\left(-4 \beta N - \lambda N^2 \left( |a| + |b| + |c| +
|a+b| + |a+c| + |b-c| \right)\right).
\ee
In the remainder, we will fix the value of $N$ and look only at
the distribution of $a,b,c$.

By symmetry, the expectation values of $a,b,c$ are all zero. This
happens because there are a number of discrete symmetries of the
quiver due to which averages vanish.  For example, for every
chiral quiver there is the anti-chiral quiver in which the
orientations of all fields are reversed.   Averaging these two
will formally give $a=b=c=0$.   Similar phenomena will always
happen whenever we consider sets of quivers with symmetries. More
structure appears once we break the symmetries and look at the
average quiver in an ensemble with some symmetry breaking
conditions imposed. Suppose for example that we impose $a>0$. This
leaves a $\mathbb Z_2$ symmetry that exchanges vertices 3 and 4.
Therefore, the expectation value of $A_{34}$ will be zero.
Symmetry considerations further show that
\be
\langle \frac{1}{2} A_{12} \rangle = \langle A_{23}\rangle =
\langle A_{24} \rangle= \langle A_{31}\rangle =\langle
A_{41}\rangle .
\ee
Furthermore, each of these expectation values is proportional to
$1/\lambda N^2$.

A boundary condition that completely breaks the symmetry is to
impose that $a\geq b \geq 0$. We can always achieve this up to a
permutation of the vertices so there is no loss of generality. The
analysis of the expectation values of the number of matter fields
in this ensemble is more tedious but can still be done explicitly.
To leading order in $\epsilon=\lambda N^2$ we
obtain\footnote{Here, by $\langle A_{ij} \rangle $ we really mean
$\langle A_{ij}-A_{ji} \rangle $.}
\bea
& & \langle A_{12}\rangle  = \frac{47}{84 \epsilon}, \quad \langle
A_{32} \rangle  = \frac{4}{21\epsilon}, \quad \langle A_{31}
\rangle = \frac{61}{588\epsilon}, \nonumber \\ & &
 \langle A_{24} \rangle =
\frac{3}{4\epsilon},\quad \langle A_{41} \rangle
=\frac{67}{147\epsilon} ,\quad \langle A_{43} \rangle =
\frac{173}{588\epsilon}.
\eea
Thus we see that after modding out the $Z_2$ symmetries of the
quiver we are able find an interesting average quiver.  Of course,
since there are only four nodes here, we do not expect any notion
of statistical typicality.  To study whether general large quivers
have some typical structure, we will have to proceed as above, by
parameterizing the space of anomaly cancelled theories  and then
imposing symmetry breaking  conditions.

\subsection{Towards dynamics}

\begin{figure}
\begin{center}
 \includegraphics{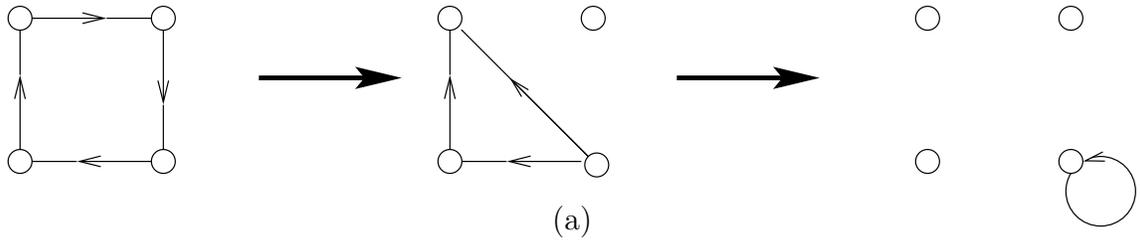} \\
 \vspace{-0.2in}
 (a) \\
\vspace{0.5in}
\includegraphics{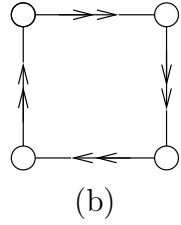} \\
(b) \\
\vspace{0.5in}
\includegraphics{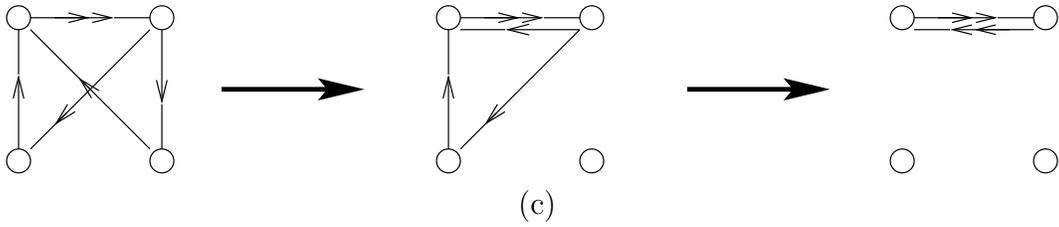} \\\
(c)  \\
\vspace{0.5in}
\includegraphics{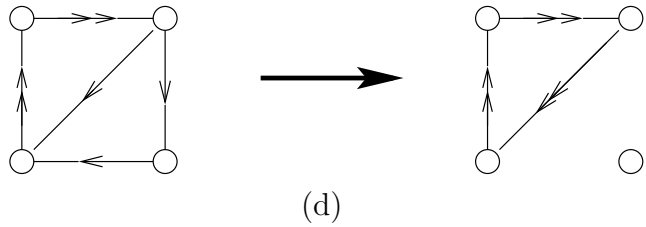} \\
(d) \vspace{0.2in} \caption{Examples of flows to the infrared in
asymptotically free, four node quiver gauge theories with equal
rank gauge groups.  The tree level superpotential is assumed to be
trivial.   (a) This quiver flows to a theory with a Coulomb branch
at low energies.   (b) This theory is a conformal field theory at
low energies. (c) Assuming the two gauge groups with $N$ flavors
have a higher scale than the groups with $2N$ flavors, this theory
flows to a theory with two confined groups, with the massless
mesons of the two groups then participating in an interacting
conformal field theory.   (d) Assume that the group that confines
has a higher dynamical scale than the other groups, and that the
confinement is on the Baryonic branch. The massless mesons of this
confining factor then interact with the rest  of the theory
leading to a flow to an interacting conformal field theory. }
\Label{fournodes}
\end{center}
\end{figure}

While we have been focusing on the structure of those field
theories in which anomalies cancel, we should also be paying
attention to dynamics. Since we are dealing with ${\cal N} = 1$
field theories, if $N_f > 3 N_c$ for any gauge group then it will
be infrared free.  If $N_f < 3 N_c$ it will be asymptotically
free. If $N_f = 3 N_c$ the one-loop beta function vanishes.  If we
distribute fields into a quiver, the bound of the total number of
fields will tend to cause the low rank gauge groups to contain
more fields. Thus they will tend to be infrared free.  What is
more, because, as we have seen above, anomaly cancellation
including high rank gauge groups tends to require lots of fields,
if a high rank group is connected to the rest of the quiver it
would tend to push groups in the quiver towards infrared freedom.
In general, studying RG flow requires us to know the
superpotential or at least to scan statistically over them.
Minimally, we should include all cubic and quartic terms in the
superpotential with O(1) coefficients mutiplied by the appropriate
scale.  (The cubic terms are classically marginal, and some
quartic terms are known to become marginal under RG flow.) Doing
such a dynamical analysis of general quiver gauge theories is
beyond the scope of this paper, but as an initial step to gain
some experience with how this works we will study some examples
without a superpotential.

\subsubsection{Four node, asymptotically free quivers}
First recall that $SU(N)$ gauge theory with $N$ flavors confines
at energies below its dynamical scale, while  $SU(N)$ theory with
$2N$ flavors flows to an interacting conformal fixed point. We
will assume that the confining $SU(N)$ theory is on the baryonic
branch.   We can then naively take a quiver and simply proceed to
allow individual gauge factors to confine, Seiberg dualize
\cite{seiberg} etc. as their dynamics becomes strong.  A cursory
analysis of four node, asymptotically free quivers (see some
examples with equal ranks $N$ in Fig.~\ref{fournodes}, constructed
from the vertices in Fig.~\ref{fivevertex}) suggests that one will
tend to get interacting conformal field theories in which the
mesons of the confining factors participate.    This suggests that
unparticles \cite{georgi} might be generic in these settings.

\subsubsection{General quiver with unequal gauge groups}

First consider the simple case of a loop of three gauge groups,
$SU(N_1) \times SU(N_2) \times SU(N_3)$ which has to cancel
anomalies by itself.  For example, this can happen if the 3-loop
is isolated within a larger quiver.   As we discussed, such
primitive 3-loops can be used to generate larger anomaly free
quiver gauge theories. To cancel anomalies, the $(12), (23), (31)$
links will generically contain $N_3, N_1, N_2$ bifundamentals
respectively.\footnote{The minimal solution to the anomaly
cancellation equations will actually be that the number of
bifundamentals connecting $i$ and $j$ is $N_k / {\rm GCD}(\{N_i,
N_j,N_k\})$ as in (\ref{cyclicanomaly}). But generically the GCD
will be $1$.} Thus for group $i$ to be asymptotically free one
will need that
\begin{equation}
3N_i > N_j N_k  ~~~~~~ i \neq j \neq k \, .
\end{equation}
Taking all the $N_i >  3$ and $N_1 < N_2 < N_3$,  it is clear that
$SU(N_3)$ is the only gauge group that has the possibility of
being asymptotically free.
%For example, to satisfy the above
%inequality, $N_1$ would have to exceed $N_2$ since $N_3$ is at
%least three, but this is false by assumption.
So for any
anomaly-free, chiral connected quiver with three nodes with ranks
at least $3$ either all three groups are infrared free, or only
the largest one is asymptotically free if it has sufficiently
large rank.

The same argument no longer works for connected quivers with more
than three gauge groups, still it is easy to see that generically
high rank gauge groups with links to smaller rank gauge groups
have a chance to be asymptotically free, whereas low rank gauge
groups connected to higher rank gauge groups tend to be IR free.

Now consider three cases for the dynamics of a quiver with unequal gauge groups.
\begin{enumerate}
\item The number of fields $K$ is very large.  If so,
it seems likely that in a randomly chosen field theory all
possible links in the quiver will be populated with some
multiplicity, although the links between low rank groups will be
enhanced.   In this circumstance our arguments suggests that the
entire  theory will be infrared free.
\item The number of fields $K$ is small.
Presumably the lowest rank gauge groups will tend to have matter
and the quiver will typically consist of several disconnected
smaller clusters that each for a connected quiver gauge theory.
The high rank gauge groups with little matter would then confine
at the appropriate dynamical scale.
\item For an intermediate number of fields the clusters
will percolate and presumably there is an interesting phase structure here.
\end{enumerate}

\section{Conclusion}

It is somewhat unsettling to attempt to make {\it statistical}
predictions for the structure of the theory describing nature
because, ever since Galileo, we have been fortunate that
observations and symmetries have constrained the possibilities
sufficiently to essentially give a unique theory describing the
phenomena under investigation.   But string theorists are in the
unprecedented situation of hoping to make predictions for the
fundamental theory up to the Planck scale given observations below
the TeV scale, subject to only very general constraints such as
consistency and in particular a consistent coupling to quantum
gravity. In such a situation, the best one can do is to predict
the likelihood of possible high energy theories, conditioned on
the known facts below the TeV scale, the known constraints, and
our best guess regarding the measure on the space of theories.
This is literally all that we can know.   While this sort of
Bayesian approach is unfamiliar in particle physics, it is much
less unusual in cosmology where one does conceive of ensembles of
possible universes or ensembles of domains with different
low-energy physics in a single universe. We would nevertheless
like to emphasize that we do not want to exclude the possibility
that consistency requirements plus experimental input will
eventually yield an (almost) unique fundamental theory, we are
merely entertaining the logical possibility that this will turn
out to not be the case.

In this paper we have used the uniform measure on specific
effective field theory landscapes, but it is not obvious that this
is the measure prescribed by string theory.   For example,
dynamics can play a role in determining the appropriate measure
because there can be transitions between vacua with different
properties.   Also, renormalization group flows can modify the
measure in the infrared as theories flow towards their fixed
points.  Given the correct measure, our analysis could be repeated
to find the typical predictions.   However, because the uniform
measure leads to typicality for some coarse-grained properties, an
alternative measure would have to concentrate on an exponentially
sparse part of the configuration space in order to change the
typical predictions of the uniform measure.

The general approach to model building suggested by these
considerations does not involve the usual desert with a high scale
GUT.   Instead it appears that one would statistically expect a
plethora of gauge factors leading to interesting structures at all
scales up to the string scale.   Amongst these gauge factors there
will be some groups with high ranks and others with low ranks. If
there is a bound on the total number of matter fields,
statistically, the higher rank groups will tend to have fewer
fundamentals (since this eats up matter).   Thus  they will tend
towards confinement at a relatively high dynamical scale if all
couplings are unified at the string scale.    On the other hand if
you have too much matter in any group it will tend to infrared
triviality.  Thus the low rank groups, if they are to have IR
dynamics, will tend to be largely decoupled from the high rank
groups.   Thus if we study the statistics of anarchic landscapes
of field theories, conditioned on having interesting low energy
dynamics, we will tend towards a structure with dynamical low rank
groups largely decoupled from a complex, interacting higher rank
sector.

The explicit examples of toy landscapes that we  studied in Sec.~3
do not have very interesting dynamics.  The matter-free case
confines.   The ring quivers that we studied in detail  are
generically infrared free  since anomaly cancellation imposes the
need for lots of matter unless the individual gauge group ranks
conspire to make the GCD in (\ref{cyclicanomaly}) large.   Thus we
see that conditioning a field theory landscape on having
interesting low energy dynamics, along with anomaly cancellation,
will be a major constraint, and is likely to significantly modify
the measure on the space of theories.    It would amusing if
curious number theoretic properties like the appearance of large
GCDs will have to be given more weight. It would also be very
interesting to explore other measures; for example the results in
\cite{timoetal,douglastaylor} suggest to weigh rank $k$ gauge
group factors with an extra factor of $1/k^2$ compared to the
anarchic measures we have been using.

\paragraph{Acknowledgments: }
We have benefited from conversations with Ofer Aharony, Michael
Douglas, Gary Gibbons, Florian Gmeiner, Dieter L\"{u}st, Juan
Maldacena, Yasunori Nomura, Carlos Nunez, Al Shapere, Tanmoy
Vachaspati, Brian Wecht, and Timo  Weigand.      We are grateful
to the organizers of the Sowers Workshop at Virginia Tech where
this work was initiated.   V.B. thanks DAMTP, Cambridge, and the
Physics Department at Swansea, and  A.N. thanks the Physics
Departments at Penn, the University of Amsterdam and the IAS for
for hospitality during this project.  VB was supported by the DOE
under grant DE-FG02-95ER40893, by the NSF collaboration grant
OISE-0443607, and as the Helen and Martin Chooljian member of the
Institute for Advanced Study. AN is supported by a STFC advanced
fellowship. JdB is partially supported by the FOM foundation.
Finally, we have enjoyed the atriums of the British Library and
the British Museum while this paper was being completed.

\end{document}